\def\cm{{\rm\thinspace cm}}

\def\ga{{\rm\thinspace gauss}}
\def\K{{\rm\thinspace K}}
\def\keV{{\rm\thinspace keV}}

\documentstyle[psfig]{mn}
\begin{document}
\hsize=6truein

\title{X-ray reflection spectra from ionized slabs}

\author[]
{\parbox[]{6.in} {R.R.~Ross$^1$, A.C.~Fabian$^2$ and  A.J~Young$^2$ \\
\footnotesize
1. Physics Department, College of the Holy Cross, Worcester, MA 01610,
USA \\
2. Institute of Astronomy, Madingley Road, Cambridge CB3 0HA \\ }}
\maketitle

\begin{abstract}
X-ray reflection spectra are an important component in the X-ray spectra
of many active galactic nuclei and Galactic black hole candidates. It is
likely that reflection takes place from highly ionized surfaces of the
accretion disc in some cases. This can lead to strong Comptonization of
the emergent iron, and other, absorption and emission features. We
present such reflection spectra here, computed in a self-consistent
manner with the method described by Ross and Fabian. In particular we
emphasise the range where the ionization parameter (the flux to density
ratio) $\xi$ is around and above $10^4$. Such spectra may be relevant to
the observed spectral features found in black hole candidates such as
Cygnus X-1 in the low/hard state.

\end{abstract}

\begin{keywords} 
accretion, accretion discs -- galaxies: active -- line: profiles -- 
radiative transfer -- X-rays: general -- X-rays: stars
\end{keywords}

\section{INTRODUCTION}
Compton reflection is an important component of the X-ray spectrum of
many compact objects (Guilbert \& Rees 1988; Lightman \& White 1989),
particularly of Active Galactic Nuclei (AGN) and Black Hole Candidate
(BHC) sources. It is produced when the primary X-ray emission from the
source strikes Thomson-thick matter which Compton scatters X-rays into
our line of sight. In many cases the accretion flow, probably a disc, is
the major scattering medium. If that matter were solely composed of
hydrogen then the reflection continuum would have the same spectral shape
as the primary emission in the X-ray band, decreasing above about 30~keV
due to Compton recoil. Other elements such as oxygen and iron, if they
are not highly ionized, can absorb the lower energy X-rays and so flatten
the reflection continuum and, in particular iron, add fluorescent line
emission (George \& Fabian 1991; Matt, Perola \& Piro 1991). Measurements
of the strength and shape of the reflection spectrum can yield the
geometry, velocity, gravitational potential depth and abundances of the
scattering medium.

As the X-ray irradiation of the medium becomes more intense, so the
matter becomes ionized. This reduces the effects of absorption,
progressively from lower to higher energies as the intensity increases
(Lightman \& White 1988; Done et al 1992; Ross \& Fabian 1993; \.Zycki et
al 1994). An important signature of ionized reflection is a strong iron
edge. Iron is a strong absorber when the gas is neutral, but absorption
both above and below the edge are so strong that the change in observed
flux across the edge is small, particularly if the primary continuum is
also observed. When much of the oxygen in the gas is completely ionized
the edge appears much stronger because the absorption below it is
reduced. Of course the energy of the iron emission line shifts up in
energy from 6.4 through 6.7 to 6.97~keV as iron is ionized, but doppler
shifts may confuse precise estimates of the observed energy. The line can
be weak when iron is ionized to Fe~{\sc xvii}--Fe~{\sc xxiii} since line
photons can be resonantly scattered and destroyed by Auger events (Ross,
Fabian
\& Brandt 1996). Moreover, as the fraction of completely stripped iron
increases the line may be significantly broadened by Comptonization and
so become less apparent in the observed spectrum (Matt, Fabian \& Ross
1996). Relativistic blurring when the reflection is from an accretion
disc can further merge the line and edge so that neither is distinct in
the final spectrum. The net result can be a weak broad absorption trough
which starts at about 7~keV. 

Unlike the reflection spectrum from relatively neutral gas, which depends
mainly upon the relative abundance of the elements and can be computed in
a straightforward manner by the Monte-Carlo technique, the spectrum when
the gas is partially ionized requires detailed numerical calculation to
obtain the temperature, ionization structure and resultant Comptonization
(Ross 1979). We have previously computed examples appropriate for simple
accretion discs around AGN and BHC, concentrating mainly on ionization
parameters $\xi=4\pi F/r^2\la 10^3$ where Auger destruction can be
important (Ross \& Fabian 1993). Since however the precise density
structure of the outer few Thomson depths of an accretion disc where the
reflection spectrum is formed is unknown (the total disc thickness may be
hundreds of Thomson depths), we consider a wider range of conditions here
and highlight the range at higher values of $\xi\sim 10^4$, the details
of which have been largely ignored in previous work.

This range of $\xi$ may be particularly relevant to the spectra of many
BHC in the low/hard state, which show only a small reflection component
(Ebisawa 1991; Ebisawa et al 1994, 1996; Gierli\'nski et al 1997;
\.Zycki, Done \& Smith 1998; Done \& \.Zycki 1998). One popular
interpretation for this is that the central parts of the disc out to
30 -- 50 gravitational radii are missing and replaced by a hot cloud
which gives no intrinsic reflection (Gierli\'nski et al 1997; \.Zycki
et al 1998; Poutanen 1998). What reflection is seen comes from the
outer disc. We point out the alternative that, in a geometry where the
primary X-rays are produced above a disc, an apparent lack of
reflection could indicate high reflection. In principle, there is no
difference in the spectrum expected from a single hot cloud and one
with a perfect mirror bisecting it. In practice, Compton reflection is
not perfect, but departures from a uniform albedo can be small in the
1--30~keV range where most spectral fitting takes place. We show here
that the small departures from uniformity due to iron features that
occur in a highly ionized disc can account for the small reflection
signature seen, without the need to illuminate the outer disc. This
solution may also be relevant to the X-ray spectra of many quasars,
which also show little evidence for reflection yet are generally
assumed to be powered by disc accretion (Nandra et al 1995, 1997).

We note that values of $\xi\sim10^4$ appear to imply an accretion rate
close to the Eddington limit for a standard accretion disc (Ross \&
Fabian 1993 show that $\xi=7\times 10^4 f^3,$ where $f$ is the Eddington
ratio). However, if there is a patchy corona above the disc (see e.g.
Stern et al 1995), rather than a continuous one, with a covering fraction
of $f_{\rm A}$ then $\xi>10^4$ requires $f>f_{\rm A}/7,$ which is only a
per cent or so when $f_{\rm A}\sim 0.1.$ Moreover, these estimates assume
a uniform density for the disc right to its surface. The outer few
Thomson depths could well have a lower density and thus mean that $\xi$
is underestimated.

In this paper we present and discuss detailed, self-consistent
computations of the temperature and ionization structure of slabs of gas
ionized by the incident radiation and of the resulting reflection
spectra. We assume both a constant density for the gas and a gaussian
fall off. The spectra will be compared with X-ray observations of BHC in
future work.

\section{Computations}

\subsection{Method}

We consider a slab of gas whose surface is illuminated by radiation
with a power-law spectrum of photon index $\Gamma$.  The radiative
transfer is calculated for the upper layer of the slab that is
responsible for producing the reflected spectrum.  In order to
concentrate solely on the effects of the illumination, no radiation is
taken to enter the treated layer from the remainder of the slab
beneath it.

The method used has been described in detail by Ross \& Fabian (1993).
The illuminating radiation is treated analytically in a `one-stream'
approximation.  The diffuse radiation that results from Compton
scattering of the incident radiation and from emission within the gas
is treated using the Fokker-Planck/diffusion method of Ross, Weaver \&
McCray (1978) modified for plane-parallel geometry.  The local
temperature and ionization state of the gas is found by solving the
equations for thermal and ionization equilibrium, so that they are
consistent with the local radiation field.  Hydrogen and helium are
assumed to be fully ionized, while the following ionization stages of
the most abundant  metals are treated: C~{\sc v--vii}, O~{\sc v--ix},
Mg~{\sc ix--xiii}, Si~{\sc xi--xv}, and Fe~{\sc xvi--xxvii}.

For a given value of $\Gamma$, the temperature and ionization state of
the outer portion of the slab is expected to depend primarily on the
value of the ionization parameter,
\begin{equation}
\xi = {4\pi F\over n_{\rm H}},
\end{equation}
where $F$ is the total illuminating flux (from $0.01-100\keV$) and
$n_{\rm H}$ is the hydrogen number density.  For uniform-density slabs,
we vary $\xi$ by changing the total illuminating flux while keeping the
hydrogen number density fixed at $n_0=10^{15}\cm^{-3}$.  This is a
typical density that might be found in an AGN accretion disc.  At the
higher density expected for a BHC accretion disc, the effects of
three-body recombination and heating due to free-free absorption cause
the temperature and ionization state to depend somewhat on $n_{\rm H}$ as
well as $\xi$ (e.g., see Ross 1979).  However, this should have very
little effect on the X-ray spectral features due to iron, which should
remain similar to those calculated here.

\subsection{Results}

Figure 1 shows the results for a uniform slab illuminated by a
$\Gamma=2$ spectrum with an ionization parameter $\xi=10^4$.
\begin{figure}
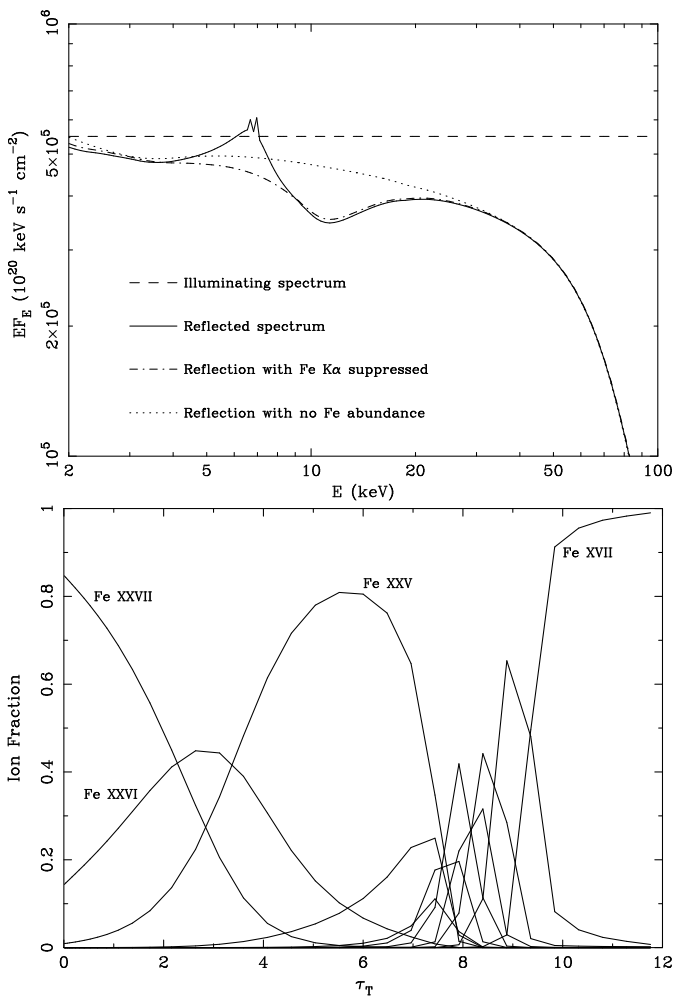
 % for Figure 1
\centerline{\psfig{figure=xi1e4ka.ps,width=0.5\textwidth,angle=270}}
\centerline{\psfig{figure=xi1e4fe.ps,width=0.5\textwidth,angle=270}}
\caption{Illumination of a uniform slab with $\Gamma=2$ and $\xi=10^4$.
 Top panel: Reflected spectrum.  Bottom panel: Iron ion fractions
 versus  Thomson depth.}
\end{figure}
The illuminating and reflected spectra are displayed as $EF_E$ versus $E$,
where $E$ is the photon energy and $F_E$ is the spectral energy flux.  The
slab is highly reflecting, with $F_E({\rm out})/F_E({\rm in})>63$ per cent
throughout the 2--20 keV spectral band.  This is because the gas is highly
ionized.  At the illuminated surface, the iron is 85 per cent fully ionized,
14per cent Fe~{\sc xxvi}, and 1 per cent Fe~{\sc xxv}.  The Fe~{\sc xxvi}
fraction peaks at 45per cent around Thomson depth $\tau_{\rm T}\sim 2.5$.
Fe~{\sc xxv} is the dominant ion for $3.5\la \tau_{\rm T}\la 7.5$.  Silicon
and magnesium (not shown in Fig.~1) are fully ionized throughout the regions
where Fe~{\sc xxv--xxvii} dominate. For $\tau_{\rm T}\ga 10$, iron ions have
all their L-shell electrons, while magnesium and silicon ions have filled
K-shells.

Despite the high ionization of the surface layers, the reflected spectrum
shows features due to iron K$\alpha$ emission and K-shell absorption.  Most
of the K$\alpha$ photons emerge in a broad Comptonized emission feature that
blends smoothly into the Compton-smeared absorption feature.  Only a small
fraction of the K$\alpha$ photons emerge in narrow Fe~{\sc xxv} and Fe~{\sc
xxvi} line cores at 6.7 and 7.0 keV, respectively; these are shown in Fig.~1
with a spectral resolution $\delta E/E\approx 2$ per cent.  Fe~{\sc xxvi}
K$\alpha$ photons are subject to resonance trapping.  Many are removed from
the narrow line core by an initial Compton scattering and then are further
Comptonized as they diffuse outward to the surface. The Fe~{\sc xxv}
intercombination line is not subject to resonance trapping, but most of
these photons are produced at such great depth that they also suffer
repeated Compton scatterings before escaping. To clarify the contribution of
the Comptonized line, Fig.~1 also shows the reflected spectrum when the iron
K$\alpha$ emission is artificially suppressed.  The K$\alpha$ photons form
an important part of the reflected continuum for $4\la E\la 9\keV$.  Also
shown in Fig.~1 is the reflected spectrum if all iron features are
suppressed by setting the iron abundance to zero.  Relative to this smooth
spectrum, the actual reflection spectrum is enhanced for $E<7.5\keV$ and
depleted for $E>7.5\keV$

For soft X-rays ($E\leq 1.5\keV$ in this case) not shown in Fig.~1,
the emergent spectral flux exceeds the incident flux due to 
bremsstrahlung emission by the hot surface layers and ``inverse
Compton'' upscattering of even softer photons.  The {\it total} flux
leaving the surface over the entire spectral range under consideration
($0.01\keV\leq E \leq 100\keV$) equals the total incident flux, as
required by the condition of thermal equilibrium.  The reflected
spectrum declines steeply above $\sim 50\keV$ because of the sharp  
cutoff assumed at $100\keV$.  Extending the illuminating
spectrum to higher energies (with  an exponential cutoff, say) would
raise this portion of the emergent spectrum via Compton downscattering
of higher energy photons, but would have little  effect on the iron
ionization structure and spectral features.

The temperature at the illuminated surface is found to be
$1.5\times 10^7\K$.  In the highly ionized gas there,
heating due to Compton downscattering of hard photons dominates 
over heating due to photoabsorption.  Therefore the temperature should 
not exceed the ``Compton temperature,'' $T_{\rm C}$, at which Compton 
heating would be balanced solely by cooling due to ``inverse Compton'' 
upscattering of soft photons.  This condition is given by
\begin{equation}
  4kT_{\rm C}\int_{E_1}^{E_2} u_E\,dE = 
  \int_{E_1}^{E_2} u_E\left(E-{21E^2\over 5m_{\rm e}c^2}\right)dE,
\end{equation}
where $u_E$ is the spectral energy density of the
radiation, while $E_1$ and $E_2$ are the lower and upper limits,
respectively, of the spectrum under consideration.  The right-hand
side of this equation includes the reduction in Compton heating due to
first-order Klein-Nishina corrections to the scattering cross section
(see Ross 1979).  Setting $u_E\propto E^{1-\Gamma}$ for the illuminating
radiation, this gives $T_{\rm C}=1.9\times 10^7\K$ for $\Gamma = 2$.
The temperature that we find at the surface is slightly lower than the 
Compton temperature due to additional cooling by bremsstrahlung emission.

This disagrees with the results of \.Zycki et al.\ (1994), who found a
surface temperature exceeding $4\times 10^7\K$ for illumination with $\xi
= 10^4$ by a spectrum that was only slightly flatter ($\Gamma = 1.9$).
The Monte Carlo calculation of \.Zycki et al.\ only treated photons in
the range $0.15\keV\leq E \leq 100\keV$ and did not include the thermal
emission by the gas itself.  This leads to underestimation of the inverse
Compton cooling rate and thus to the high surface temperature (and the
steep temperature gradient) that they found.

Figure 2 shows a series of reflection spectra under similar conditions
with ionization parameters ranging from $\xi=30$ to $\xi=10^5$.  
\begin{figure*} % for Figure 2
\centerline{\psfig{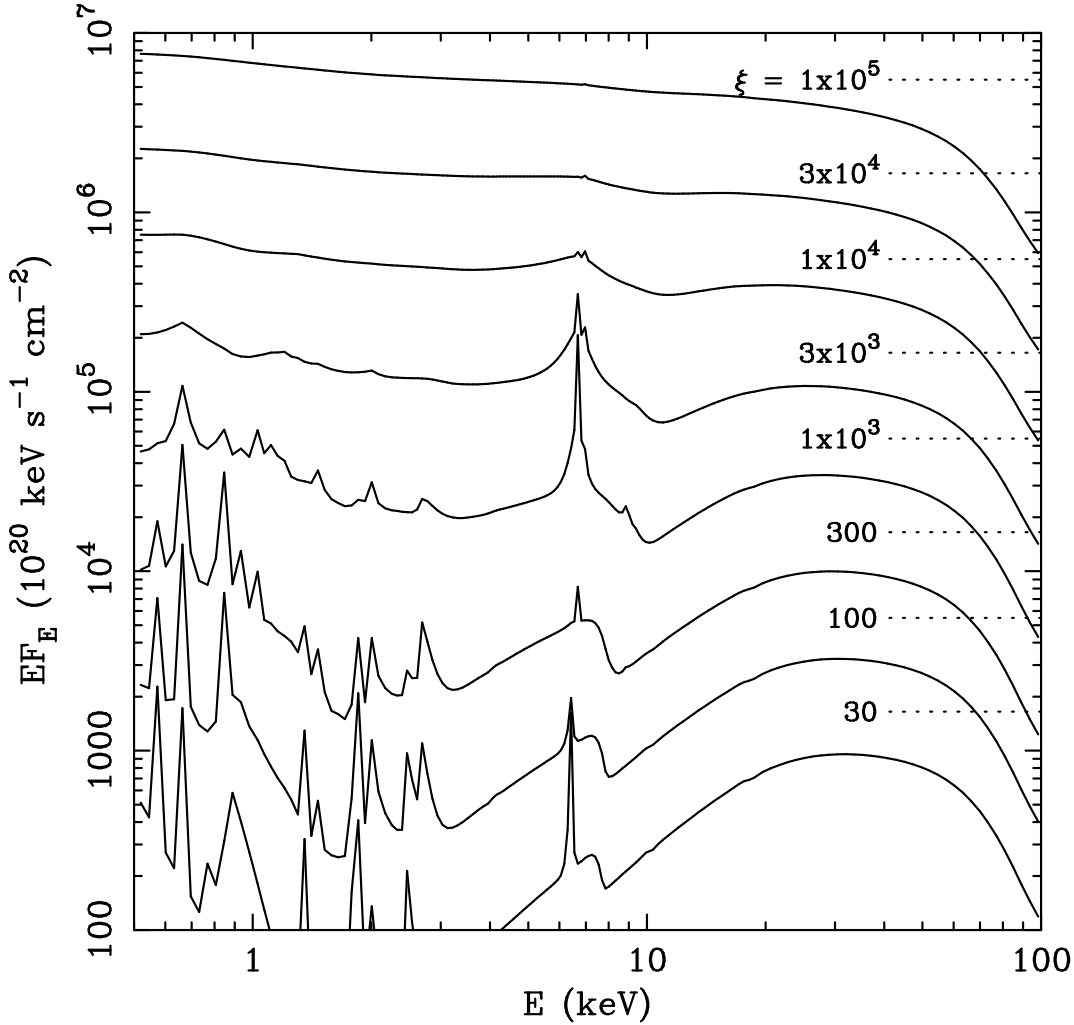}}
\caption{Reflection spectra for illumination of a uniform slab with
 $\Gamma=2$ and various values of the ionization parameter.  The
 number above each curve indicates the corresponding $\xi$ value, and
 the dotted line to the right of that number indicates the constant
 value of  $EF_E$ for that illumination.}
\end{figure*}
The model with the highest ionization parameter, $\xi=10^5$, is an
excellent reflector and exhibits negligibly small spectral features due
to  iron.  This is because the surface layer is fully ionized to great
depth,  with Fe~{\sc xxvi} not becoming dominant until $\tau_{\rm
T}\approx 8$.   The Compton reflection produces a slight steepening in
the reflected  spectrum compared to the illumination; the reflected
component mimics a power law with $\Gamma=2.14$ in the 2--20 keV band.
The temperature at the illuminated surface is found to be
$1.9\times 10^7\K$, the full Compton temperature for a $\Gamma = 2$ 
power law spectrum.

When the illuminating flux is reduced so that $\xi=3\times 10^4$, the
K$\alpha$ emission and K-shell absorption features due to iron begin
to  become apparent.  Fully-ionized iron dominates for $\tau_{\rm
T}\la 5$,  so the emission and absorption features are weak and are
highly broadened  by Compton scattering.  When $\xi$ is reduced to
$10^4$, the broad spectral features due to iron become more important.
This model has already been discussed in detail.

With $\xi$ reduced to $3\times 10^3$, less than half of the iron is
fully ionized at the illuminated surface, and Fe~{\sc xxv} becomes
dominant at $\tau_{\rm T} \approx 1$.  Now an important {\it narrow}
emission line due to Fe~{\sc xxv} can be seen in addition to a
Compton-broadened emission feature in Fig.~2.  This is primarily due
to emission of the intercombination line following recombination to
excited states.  When $\xi$ is further reduced to $10^3$, Fe~{\sc xxv}
dominates at the illuminated surface, and the narrow Fe~{\sc xxv}
K$\alpha$ line is quite strong.  The tiny emission feature just
above 8.8~keV is due to radiative recombination directly to the
ground level of Fe~{\sc xxv}.  Similarly, Si~{\sc xiv} produces two
emission features: one at $2.01\keV$ due to K$\alpha$ emission and 
the other just above $2.67\keV$ due to radiative recombination. (See 
\.Zycki et al.\ 1994 for other examples of radiative recombination 
emission features.) The exact strength of these and other low-energy 
spectral features could be affected by absorption by elements such as 
nitrogen and neon which we do not include in our calculations, and we 
do not calculate the features due to sulfur.

The narrow Fe K$\alpha$ emission line is suppressed in the models with
$\xi = 300$ and $\xi = 100$.  This is because ions in the range
Fe~{\sc xvii}--Fe~{\sc xxiii} dominate at the illuminated surface, and
their K$\alpha$ photons are assumed to be destroyed by the Auger
effect during resonance trapping (see Ross \& Fabian 1993; \.Zycki \& 
Czerny 1994; Ross et al. 1996).   The narrow line seen when $\xi = 300$ 
is due to a small amount of  Fe~{\sc xxv} near the surface.  On the 
other hand, the narrow line seen  when $\xi = 100$ is due to 
Fe~{\sc xvi}, the least ionized species that we treat, which then 
dominates for $\tau_{\rm T}\ga 0.5$.  Finally, for  $\xi = 30$ the 
reflection is similar to that of a cold, neutral slab, and the narrow 
emission line at 6.4~keV is strong.

The ionization structure in the outer layers of the illuminated slab
depends on the spectral form of the illumination as well as on the
ratio of total flux to gas density expressed by the parameter $\xi$.
\begin{figure}
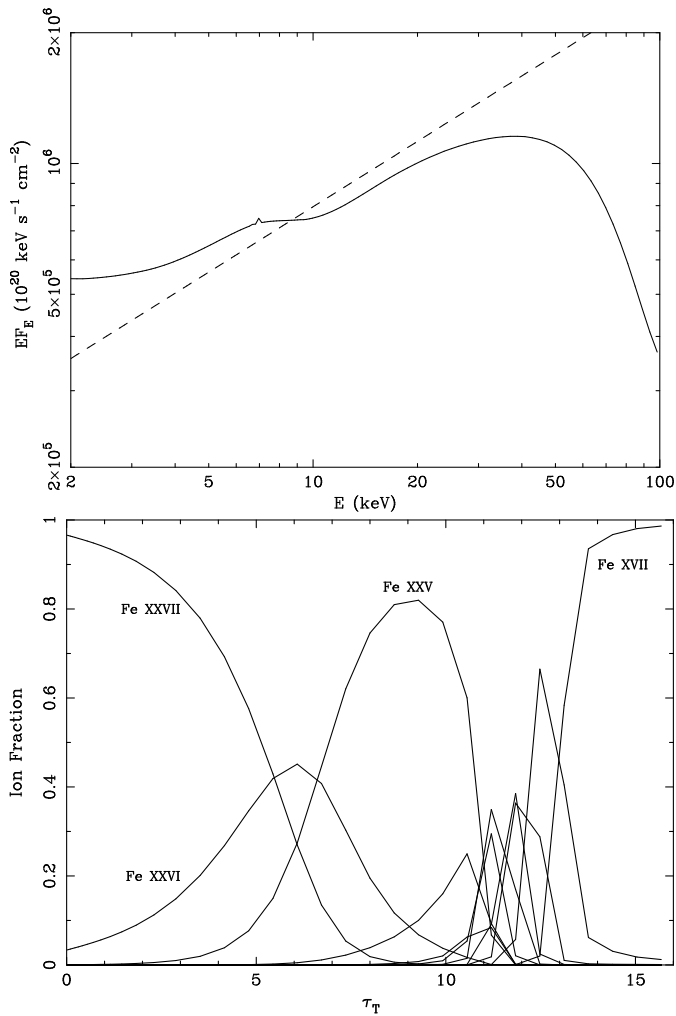
 % for Figure 3
\centerline{\psfig{figure=xi1e4g15.ps,width=0.5\textwidth,angle=270}}
\centerline{\psfig{figure=xi1e4g15fe.ps,width=0.5\textwidth,angle=270}}
\caption{Illumination of a uniform slab with $\xi=10^4$ for a $\Gamma=1.5$
 power law.  Top panel: Reflected spectrum (solid curve) and
 illumination (dashed line).  Bottom panel: Iron ion fractions versus
 Thomson depth.}
\end{figure}
Figure~3 shows the results with $\xi=10^4$ again, but when the
illuminating spectrum is a flatter power law with $\Gamma=1.5$.  Now a
greater fraction of the illuminating photons lie in the 9--20 keV
range that is so important in producing fully photoionized iron.  As a
result, Fe~{\sc xxvii} dominates to greater depth ($\tau_{\rm T}\la
5.5$) than when $\Gamma=2$, and the Compton-broadened emission and
absorption features due to iron are not as strong.

One of the uncertainties in modelling Compton reflection is the
density  structure of the illuminated gas.  In the models presented
above, the gas density has been assumed to be uniform with depth.
This is the case, for example, in the standard Shakura \& Sunyaev
(1973) theory of accretion discs (also see Laor \& Netzer 1989).
However, this probably is not  realistic even for a bare accretion
disc (e.g., see Shakura, Sunyaev \&  Zilitinkevich 1978; Shimura \&
Takahara 1993), and it certainly cannot be the case when the surface
has strong external illumination.  The heating  of the outermost
layers by the impinging radiation should result in a  decrease in
density there.

In order to see the general effect of such a decrease in density, let
us arbitrarily assume that a constant-density slab (with $n_{\rm
H}=n_0$)  is topped by a ``boundary layer'' in which the density
decreases with  height in a gaussian manner,
\begin{equation}
n_{\rm H}(z) = n_0 \exp\left(-\frac{z^2}{h^2}\right).
\end{equation}
Here $z$ is the height above the base of the boundary layer, and $h$
is  its characteristic thickness.  Since the illuminating radiation is
expected to have an important effect down to a Thomson depth of a few,
we let the boundary layer have a total Thomson depth
\begin{equation}
\tau_{\rm T} = 1.2 n_0 \sigma_{\rm T} h \frac{\sqrt{\pi}}{2} = 5,
\end{equation}
where the free electron density is assumed to be $n_{\rm e} = 1.2
n_{\rm H}$.
\begin{figure} % for Figure 4
\centerline{\psfig{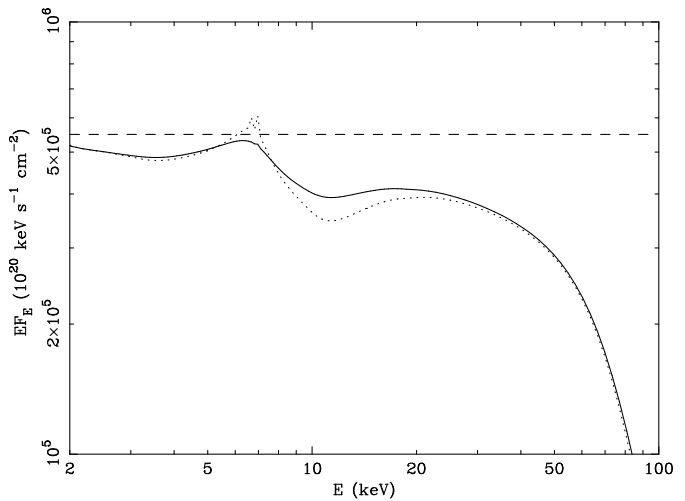}}
\caption{Illumination with $\Gamma=2$ and $\xi_0=10^4$ when there
 is a gaussian dropoff in density at the top of the slab with a
 Thomson depth $\tau_{\rm T}=5$.  The reflected spectrum and the
 incident illumination are shown by the solid and dashed curves,
 respectively.  Also shown for comparison is the reflected spectrum
 (dotted curve) for  uniform density from Fig.~1}
\end{figure}
Figure~4 shows the result when such a gas is illuminated by a $\Gamma=2$
power law with a flux that would yield an ionization parameter $\xi_0=4\pi
F/n_0=10^4$ for gas at the base density.  Since the gas density is lower in
the boundary layer, the effective ionization parameter is higher there.  As
a result, iron is 88 per cent fully ionized at Thomson depth $\tau_{\rm
T}=1$, compared to only 69 per cent fully ionized for the uniform-density
slab shown in Fig.~1.  In fact, Fe~{\sc xxvii} dominates all the way down to
$\tau_{\rm T}\approx 3.5$.  The broad, Comptonized, iron K$\alpha$ emission
and K-shell absorption features in the reflected spectrum are not as strong
as for the uniform-density slab.  Of particular importance is the fact that
the narrow K$\alpha$ line cores, which were already weak in the
uniform-density case, are now almost totally suppressed.

\section{Discussion}

For X-ray illumination with $\xi\sim 10^4$ or higher, the features in
the reflected spectrum due to iron K$\alpha$ emission and K-shell
absorption are weak and are smeared out by Compton scattering.  Any
narrow Fe K$\alpha$ line cores are extremely weak.  These effects are
further enhanced when the illuminating spectrum is harder (flatter) or
when there is a dropoff in gas density due to the heating by the
external illumination.

Such effects may come into play in the formation of X-ray spectra of
Black Hole Candidates.  In the past, {\it Ginga} spectra of BHCs have
been  interpreted as exhibiting broad iron absorption features
(``smeared  edges'') with very weak Fe K$\alpha$ lines (Ebisawa 1991;
Tanaka 1991; Ebisawa et al.\ 1994).  This led to the suggestion that
the line is suppressed by resonant Auger destruction in Fe~{\sc
xvii--xxiii}  (Ross \& Fabian 1993; Ueda, Ebisawa \& Done 1994; Ross
et al.\ 1996).

However, recent spectral studies of {\it Ginga}, {\it EXOSAT}, and {\it
ASCA} BHC in the low/hard state have found broad Fe K$\alpha$ emission
features as well as K-shell absorption features (\.Zycki, Done \& Smith
1997; \.Zycki et al.\ 1998; Done \& \.Zycki 1998).  These studies have
treated the broadening of the features as being due to relativistic
smearing by the reflecting accretion disc.  The $\xi$ values derived for
the illumination have been very low, and the weakness of the iron
features has been taken to imply that the disc subtends a solid angle of
illuminating radiation considerably smaller than $2\pi$.  (Done \&
\.Zycki conclude that the broadening of the disc K$\alpha$ line makes
it difficult to detect with {\it ASCA}, so the narrow line found by
Ebisawa et al.\ 1996 comes from the companion star.)  This has led to
the conclusion that the optically thick accretion disc only extends
inward to a few tens of gravitational radii.

As an alternative, the broad and weak iron features could be due to
illumination of the disc with $\xi\ga 10^4$.
\begin{figure}
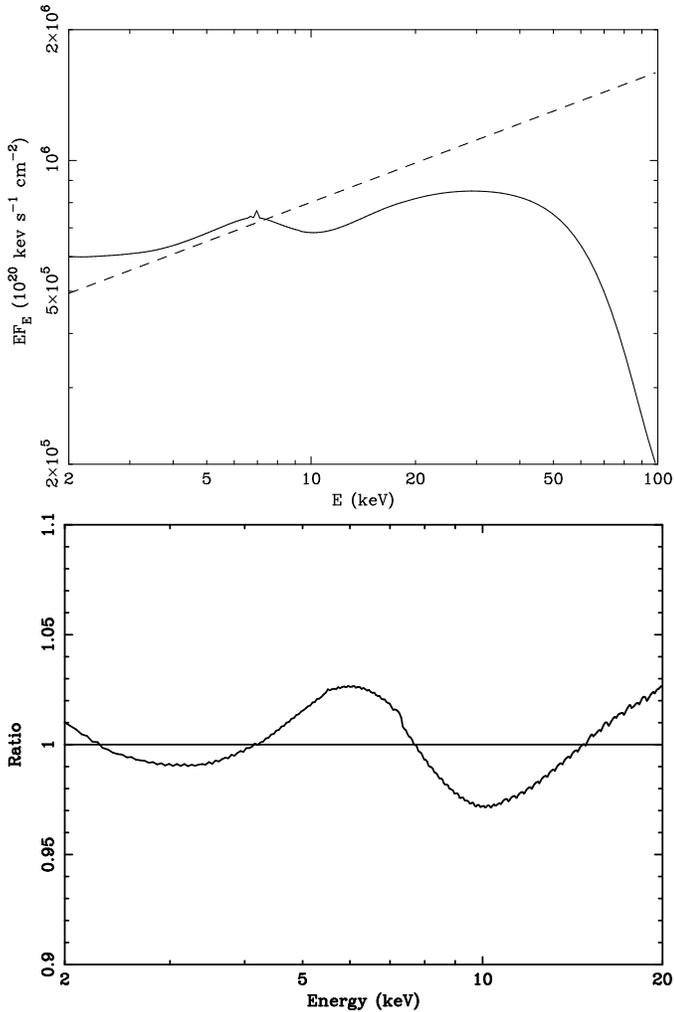
 % for Figure 5
\centerline{\psfig{figure=xi1e4g17.ps,width=0.5\textwidth,angle=270}}
\centerline{\psfig{figure=ratio.ps,width=0.5\textwidth,angle=270}}
\caption{Illumination of a uniform slab with $\xi=10^4$ for a $\Gamma=1.7$
 power law.  Top panel: Illumination (dashed line) and reflected
 spectrum (solid curve).  Bottom panel: Ratio of total spectrum
 (incident plus  reflected) to best-fit power law model after taking
 into account smearing by relativistic effects for emission by an
 annulus at $7R_S$ around a black hole.}
\end{figure}
Figure~5 shows the incident and reflected spectra for illumination
with  $\xi = 10^4$ and a power law index $\Gamma = 1.7$.  If the
primary  X-ray emission is from a corona immediately above the
accretion disc,  the reprocessor subtends a full $\Omega = 2\pi$, and
the total observed  spectrum is the sum of the illuminating flux (due
to the half of the  radiation emitted in the outward direction) and
the reflected flux.   Fig.~5 shows the result after relativistic
smearing if this total flux is assumed to originate from a narrow
annulus at radius $7R_S$ inclined at $30^\circ$ around a Kerr black
hole.  Over the 2--20 keV range, the best-fitting power law now has a
slope of $\Gamma = 1.79$.  The ratio of the observed spectrum to the
best-fit power law model shows broad emission and absorption features
markedly similar to those found by Done \& \.Zycki (1998) for the 
{\it EXOSAT} spectrum of Cyg X-1.
There are several features to note here: Comptonization of the edge means
that it begins around 6~keV (dash-dot curve in Fig.~1) and now resembles 
more a symmetrical trough than a convential photelectric absorption edge;
the Comptonized and smeared line emission fills in the lower energy part 
of this trough so that the whole feature mimics an edge at about 7~keV.

\begin{figure}
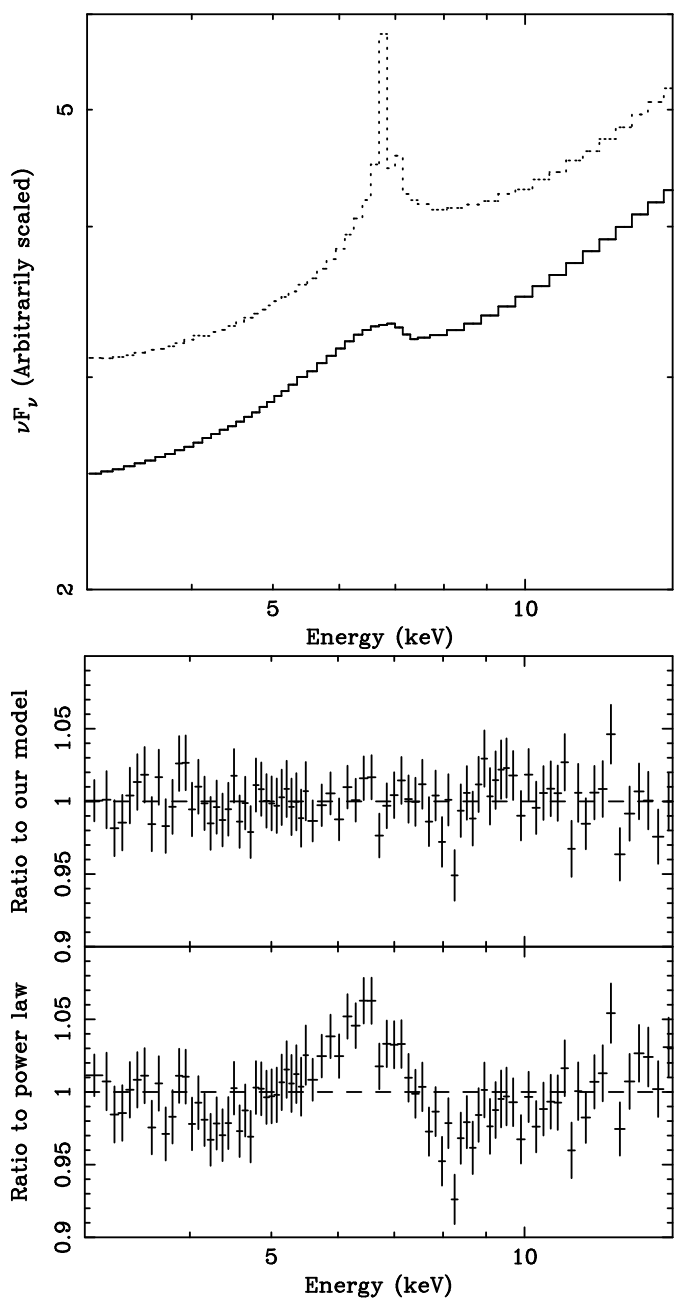
 % for Figure 6
\centerline{\psfig{figure=our_model_sum.ps,width=0.5\textwidth,angle=270}}
\centerline{\psfig{figure=ratio_plot.ps,width=0.5\textwidth,angle=270}}
\caption{Illumination of uniform slab with $\xi=7400$, $\Gamma=1.5$ and
twice Solar iron abundance. Upper panel: Unblurred (dotted line) and
relativistically blurred (solid line) illuminating plus reflected
spectra. Centre panel: Ratio of data to our model (the blurred one from
the upper panel). Lower panel: Ratio of data to simple power law.}
\end{figure}

Encouraged by the similarity bewteen our model and the observed spectra
shown by Done \& \.Zycki (1999), we have fitted a grid of our models to
the brightest of the archival {\it EXOSAT} spectra. The models are
relativistically blurred using the Kerr metric kernel of Laor (1990). The
best fit over the 3--15~keV range indicates $\xi=7400$ and $\Gamma=1.5$
for an iron abundance twice the Solar value (Fig. 6, upper and centre
panels) from a disc inclined at less than 30 deg and extending from
$1.235m$ to $100m$, with surface emissivity varying as $({\rm
radius})^{-3}$. In the lower panel we show the ratio of the data to the
best-fitting power law model, which strongly resembles the similar plot
by Done \& \.Zycki (1999) for all the {\it EXOSAT} data.

It is interesting to compare the result of our calculation with that 
given by the PEXRIV code (Magdziarz \& Zdziarski 1995) in the XSPEC 
package for a value of $\xi=5000$, which is within the range allowed by 
that code (we correct for the different energy ranges used to define
$\xi$ in our codes). This is shown in Figure~7.
\begin{figure} % for Figure 7
\centerline{\psfig{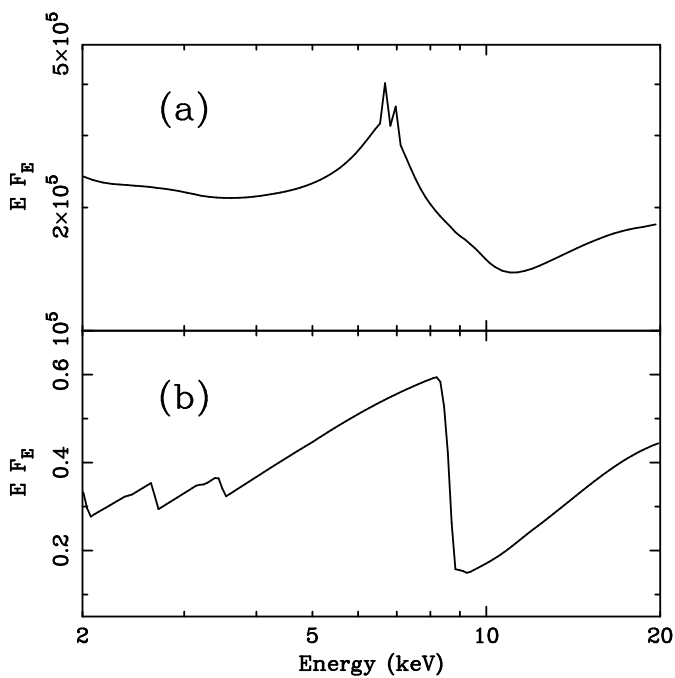}}
\caption{The reflection component  from the present work for $\xi=5000$
(a) and from the PEXRIV model (Magdziarz \& Zdziarski 1995) in XSPEC with
a corresponding $\xi=4500$ (when the different energy range is corrected
for) and disc temperature of $10^6$ K (b). The PEXRIV spectrum is just
for the continuum, whereas spectrum (a) includes the iron emission lines,
the level of contribution of which can be assessed from Fig.~1. Note from
our spectra shown here, and in Fig.~2, that we never obtain a strong edge
dropping above 8~keV, as seen in the PEXRIV model.}
\end{figure}
There is clearly a large difference, particularly in the position of the 
iron edge (the PEXRIV code does not predict the iron line properties, 
which must be added separately). The PEXRIV code does not take account 
of the Comptonization of the features in the outer, most highly ionized 
layers. We advise against the use of this code for high values of $\xi$ 
and for the use of our approach, or that of \.Zycki et al (1994; see 
also B\"ottcher, Liang \& Smith 1998).

A detailed model for an ionized accretion disc will necessarily require a
range of $\xi$ to be present, both from different radii, and at different
distances from the ionizing source if the corona is patchy. The results
shown in this paper will be useful as a guide to situations dominated by
highly ionized matter, for example where the size of separate patches of
a corona exceed their height above the disc. We intend to pursue a
detailed comparison with observed spectra in future work. 

\section*{ACKNOWLEDGEMENTS} 
RRR, ACF and AJY thank the College of the Holy Cross, the Royal Society
and PPARC, respectively, for support.


\begin{thebibliography}{}
\bibitem[]{} B\"ottcher  M., Liang E. P., Smith I. A., 1998, A\&A 
 submitted, astro-ph 9806296
\bibitem[]{} Done C., Mulchaey J. S., Mushotzky R. F., Arnaud K. A., 1992,
 ApJ, 395, 275 
\bibitem[]{} Done C., \.Zycki P. T., 1998, preprint
\bibitem[]{} Ebisawa K., 1991, PhD thesis, Univ.\ of Tokyo
\bibitem[]{} Ebisawa K. et al., 1994, PASJ, 46, 375
\bibitem[]{} Ebisawa K., Ueda Y., Inoue H., Tanaka Y., White N. E., 1996,
 ApJ, 467, 419
\bibitem[]{} George I. M., Fabian A. C., 1991, MNRAS, 249, 352
\bibitem[]{} Gierli\'nski M., Zdziarski A. A., Done C., Johnson W. N., 
 Ebisawa, K., Ueda Y., Haardt F., Phlips B.F., 1997, MNRAS 288, 958
\bibitem[]{} Guilbert P. W., Rees M. J., 1988, MNRAS, 233, 475
\bibitem[]{} Laor A., Netzer H., 1989, MNRAS, 238, 897
\bibitem[]{} Lightman A. P., White T. R., 1988, ApJ, 335, 57
\bibitem[]{} Magdziarz P., Zdziarski A. A., 1995, MNRAS, 273, 837
\bibitem[]{} Matt G., Fabian A. C., Ross R. R., 1996, MNRAS, 278, 1111
\bibitem[]{} Matt G., Perola G. C., Piro L., 1991, A\&A, 247, 25
\bibitem[]{} Nandra K., Fabian A. C., Brandt W. N., Kunieda H., 
 Matsuoka M., Mihara T., Ogasaka Y., Terashima Y., 1995, MNRAS, 276, 1
\bibitem[]{} Nandra K., George I. M., Mushotzky R. F., Turner T. J., 
 Yaqoob T., 1997, ApJ, 488, L91
\bibitem[]{} Poutanen J., preprint, astro-ph 9805025
\bibitem[]{} Ross R. R., 1979, ApJ, 233, 334
\bibitem[]{} Ross R. R., Fabian A. C., 1993, MNRAS, 261, 74
\bibitem[]{} Ross R. R., Fabian A. C., Brandt W. N., 1996, MNRAS, 278, 1082
\bibitem[]{} Ross R. R., Weaver R., McCray R., 1978, ApJ, 219, 292
\bibitem[]{} Shakura N. I., Sunyaev R. A., 1973, A\&A, 24, 337 
\bibitem[]{} Shakura N. I., Sunyaev R. A., Zilitinkevich S. S., 1978, 
 A\&A, 62, 179
\bibitem[]{} Shimura T., Takahara F., 1993, ApJ, 419, 78
\bibitem[]{} Stern B. E., Poutanen J., Svensson R., Sikora M., 
 Begelman, M. C., 1995, ApJ, 449, L13
\bibitem[]{} Tanaka Y., 1991, in Treves A., Perrola G. C., Stella L.,
 eds, Lecture Notes in Physics 385, Iron Line Diagnostics in X-ray 
 Sources. Springer, Berlin, p.\ 98
\bibitem[]{} Ueda Y., Ebisawa K., Done C., 1994, PASJ, 46, 107
\bibitem[]{} \.Zycki P. T., Czerny, B., 1994, MNRAS, 266, 653
\bibitem[]{} \.Zycki P. T., Done C., Smith D. A., 1997, ApJ, 488, L113
\bibitem[]{} \.Zycki P. T., Done C., Smith D. A., 1998, ApJ, 496, L25
\bibitem[]{} \.Zycki P. T., Krolik J. H., Zdziarski A. A., Kallman T. R.,
 1994, ApJ, 437, 597

\end{thebibliography}
\end{document}